\documentclass[appendixfloats]{emulateapj}
\usepackage{longtable}
\usepackage{graphicx}
\usepackage{graphics}
\usepackage{color}
\usepackage{epsfig}
\usepackage{rotating}
\usepackage{subfigure}
\usepackage{float}
\usepackage{url}
\usepackage{hyperref}
\usepackage{breakurl}
\usepackage{scalerel}
\usepackage[titletoc,title]{appendix}

\shortauthors{J. T. Li et al.}
\shorttitle{Molecular Gas NGC~5908}

\begin{document}

\title{Detection of non-thermal hard X-ray emission from the ``Fermi bubble'' in an external galaxy}

\author{Jiang-Tao Li\altaffilmark{1}, Edmund Hodges-Kluck\altaffilmark{2,3}, Yelena Stein\altaffilmark{4,5}, Joel N. Bregman\altaffilmark{1}, Judith A. Irwin\altaffilmark{6}, Ralf-J\"{u}rgen Dettmar\altaffilmark{7}} 

\altaffiltext{1}{Department of Astronomy, University of Michigan, 311 West Hall, 1085 S. University Ave, Ann Arbor, MI, 48109-1107, U.S.A.}

\altaffiltext{2}{Department of Astronomy, University of Maryland, College Park, MD 20742, U.S.A.}

\altaffiltext{3}{X-ray Astrophysics Laboratory, Astrophysics Science Division, NASA Goddard Space Flight Center, Greenbelt, MD 20771, U.S.A.}

\altaffiltext{4}{Observatoire astronomique de Strasbourg, Universit\'{e} de Strasbourg, CNRS, UMR 7550, 11 rue de l'Universit\'{e}, 67000 Strasbourg, France}

\altaffiltext{5}{Astronomisches Institut (AIRUB), Ruhr-Universit\"{a}t Bochum, Universit\"{a}tsstrasse 150, 44801 Bochum, Germany}

\altaffiltext{6}{Department of Physics, Engineering Physics, \& Astronomy, Queen’s University, Kingston, Ontario, Canada, K7L 3N6}

\altaffiltext{7}{Ruhr-Universit\"{a}t Bochum, Fakult\"{a}t f\"{u}r Physik und Astronomie, Astronomisches Institut, 44780 Bochum, Germny}

%\keywords{}

\nonumber

\begin{abstract}
We report new \emph{Chandra} hard X-ray ($>2\rm~keV$) and \emph{JVLA} C-band observations of the nuclear superbubble of NGC~3079, an analog of the ``Fermi bubble'' in our Milky Way. We detect extended hard X-ray emission on the SW side of the galactic nucleus with coherent multi-wavelength features in radio, H$\alpha$, and soft X-ray. The hard X-ray feature has a cone shape with possibly a weak cap, forming a bubble-like structure with a diameter of $\sim1.1\rm~kpc$. A similar extended feature, however, is not detected on the NE side, which is brighter in all other wavelengths such as radio, H$\alpha$, and soft X-ray. Scattered photons from the nuclear region or other nearby point-like X-ray bright sources, inverse Compton emission from cosmic ray electrons via interaction with the cosmic microwave background, or any individually faint stellar X-ray source populations, cannot explain the extended hard X-ray emission on the SW side and the strongly NE/SW asymmetry. A synchrotron emission model, plus a thermal component accounting for the excess at $\sim1\rm~keV$, can well characterize the broadband radio/hard X-ray spectra. The broadband synchrotron spectra do not show any significant cutoff, and even possibly slightly flatten at higher energy. This rules out a loss-limited scenario in the acceleration of the cosmic ray electrons in or around this superbubble. As the first detection of kpc-scale extended hard X-ray emission associated with a galactic nuclear superbubble, the spatial and spectral properties of the multi-wavelength emissions indicate that the cosmic ray leptons responsible for the broad-band synchrotron emission from the SW bubble must be accelerated in situ, instead of transported from the nuclear region of the galaxy.
\end{abstract}

\section{Introduction}\label{sec:Introduction}

Superbubbles in the nuclear regions of galaxies could be produced either by nuclear starbursts or by AGN. A well known example of such a galactic nuclear superbubble is the ``Fermi bubble'' in our Milky Way (MW; \citealt{Su10}) and its multi-wavelength counterparts (e.g., \citealt{BlandHawthorn03,Finkbeiner04}), for which various models based on star formation or AGN feedback have been proposed (e.g., \citealt{Guo12,Crocker15,Narayanan17}). Galactic nuclear superbubbles are also potential factories of high energy cosmic rays (CRs), especially those above the ``knee'' of the CR energy spectrum (at $\sim10^{15}\rm~eV$) which cannot be accelerated within individual supernova (SN) remnants (SNRs, e.g., \citealt{Kotera11} and references therein).

CRs could produce broadband non-thermal emission in various environments. In particular, the synchrotron emission produced by CR leptons in magnetic fields is often a dominant component of the non-thermal spectrum from radio (with frequency $\sim\rm GHz$ or lower) to hard X-ray ($\gtrsim2\rm~keV$) in highly magnetized environments. This broadband synchrotron emission has been commonly detected in young or middle-aged SNRs when the shock interacts with the interstellar medium (ISM) (e.g., \citealt{Koyama95,Gaensler98,Li15}), but the galactic scale synchrotron emission has only been firmly detected in the radio band (e.g., \citealt{Irwin12a,Wiegert15}). There are many detections of the hard X-ray emission on small scales from the nuclear region of starburst galaxies or the Galactic ridge, but the emission is dominated by stellar sources which are difficult to quantitatively remove (e.g., \citealt{Strickland07,Revnivtsev09}). \citet{HodgesKluck18} found a large-scale diffuse non-thermal hard X-ray excess in the halo of NGC~891, but the nature of this emission is unclear. The lack of detection of galactic scale synchrotron hard X-ray emission may be caused by either a steeper leptonic CR energy spectrum of galactic scale CR sources (e.g., nuclear starburst regions, compared to individual SNRs, e.g., \citealt{Li16,Li18}), or an efficient synchrotron radiative cooling of high energy CR electrons (e.g., \citealt{Lacki13}), which awaits further observational test. Furthermore, energetic galactic scale superwinds have been revealed in many starburst galaxies (e.g., \citealt{Strickland02}), some with clear radio relics (e.g., \citealt{Li08}), possibly indicating the interaction of the galactic outflow with the circum- or inter-galactic medium (CGM or IGM). However, it is still not clear if such winds could efficiently accelerate CRs or just compress the circum-galactic magnetic field and enhance the local synchrotron radio emission. Observing synchrotron hard X-ray emission at a much larger scale (e.g., $\gtrsim\rm kpc$) than individual SNRs will help us to better understand these problems.

We herein present new \emph{Chandra} X-ray and the Karl G. Jansky Very Large Array (\emph{JVLA}) radio observations of NGC~3079 (assuming a distance of $d=20.6\rm~Mpc$, $1^{\prime\prime}\approx100\rm~pc$, \citealt{Tully13}), which hosts a $\sim1.3\rm~kpc$-diameter nuclear bubble bright in radio, H$\alpha$, and soft X-ray \citep{Duric88,Veilleux94,Cecil02} (a slightly smaller bubble is also visible on the opposite side; Fig.~\ref{fig:superbubbleimg}a). The present paper is primarily aimed at studying the broadband non-thermal emission, while the thermal soft X-ray emission will be discussed elsewhere. In the following sections, we will present reduction and analysis of the X-ray and radio data in \S\ref{sec:dataanalysis}, and then discuss the results and summarize our major conclusions in \S\ref{sec:discussion}. Errors are quoted at the 1~$\sigma$ level throughout the paper.

\section{Data reduction and analysis}\label{sec:dataanalysis}

\subsection{X-ray and radio data}\label{subsec:XrayRadioData}

\begin{figure*}
\begin{center}
\epsfig{figure=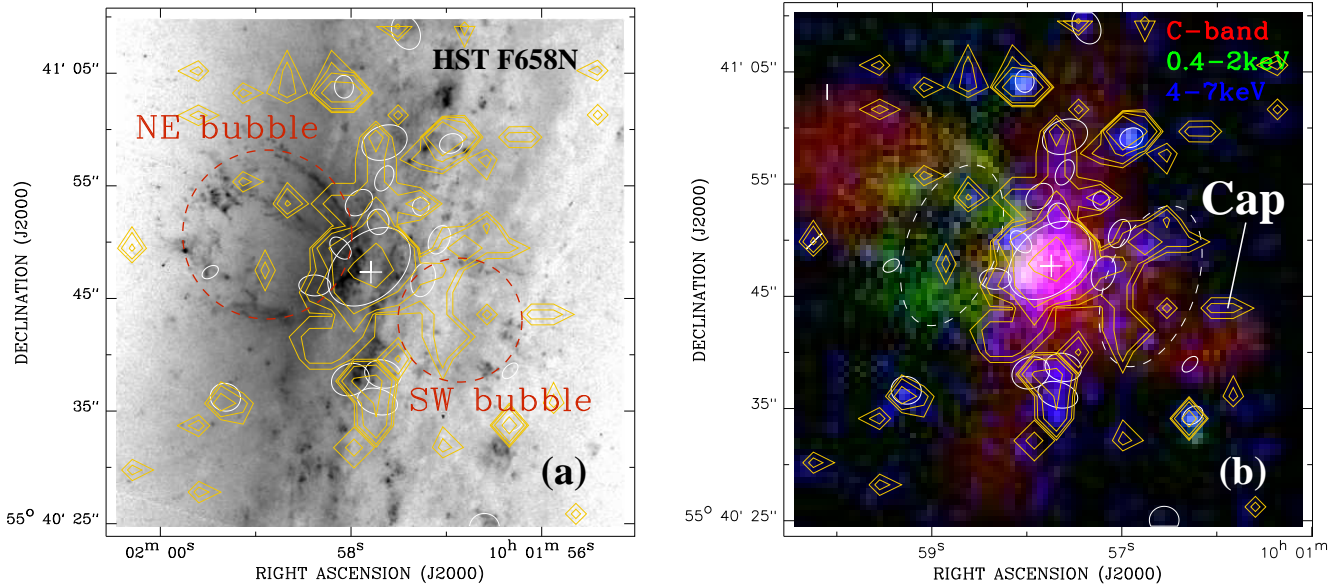,width=1.0\textwidth,angle=0, clip=}
\caption{Multi-wavelength images of the central $0.75^\prime\times0.75^\prime$ region of NGC~3079. (a) \emph{HST}/F658N (H$\alpha$) image clearly shows the bright NE bubble with a diameter of $\sim15^{\prime\prime}\sim1.5\rm~kpc$ and the fainter and smaller SW bubble with a diameter of $\sim11^{\prime\prime}\sim1.1\rm~kpc$, both outlined with dashed red circles. Yellow contours are the \emph{Chandra} 4-7~keV image shown in panel (b). Small white ellipses are the point-like X-ray sources detected with the 2-7~keV \emph{Chandra} image. The white plus marks the location of the nucleus of NGC~3079. (b) Radio and X-ray tri-color image of the same area as panel~(a): Red: \emph{JVLA} C-array C-band image; Green: \emph{Chandra} 0.4-2~keV image; Blue: \emph{Chandra} 4-7~keV image binned with a factor of 2 to highlight the faint extended features. The contours shows the smoothed unbinned 4-7~keV image at levels of 2, 3, 4, 5, 10$\times10^{-9}\rm~photons~s^{-1}~cm^{-2}$. The small white ellipse are the same as in panel~(a). The two large white dashed ellipse are the spectral extraction regions which have the same shape and area, and are symmetric to the nucleus of the galaxy. We also mark the rough location of the ``Cap'' of the SW bubble, which is visible in radio continuum but just marginally detected in hard X-ray.}\label{fig:superbubbleimg}
\end{center}
\end{figure*}

The \emph{Chandra} data used in this project are summarized in Table~\ref{table:ChandraData}, with a total effective exposure time of 128.9~ks. All the data are used to create the images, while three observations (except for that with ObsID 7851) are used to extract spectra. Data has been reprocessed using the CIAO (version 4.10, with CALDB version 4.7.8) tool \emph{chandra\_repro}, and exposure corrected images are created using \emph{merge\_obs}. We also perform point source detection in hard X-ray (2-7~keV) using \emph{wavdetect} with the wavelet radii set to 1, 2, 4, 8 pixels, in order to detect sources with a variety of fluxes (Fig.~\ref{fig:superbubbleimg}). The diffuse X-ray spectra are extracted using \emph{specextract} from two regions with the same shape and area, and are symmetric to the nucleus of the galaxy for comparison (Fig.~\ref{fig:superbubbleimg}b). A local background is also extracted from a blank region on the same CCD chip. The spectra are further regrouped to have at least 10~counts in each bin. 

\begin{table}
\begin{center}
\caption{Chandra observations of the nuclear superbubble of NGC~3079. The two deepest observations are taken in 2018 after the significant degrading of ACIS.}
\footnotesize
\tabcolsep=10pt
\begin{tabular}{lcccccccccccccc}
\hline
ObsID & Start Date & $t_{\rm exp}$ (ks) & PI \\
\hline
2038 & 2001-03-07 & 26.58 & Strickland \\
7851 & 2006-12-27 & 4.68   & Mathur \\
19307 & 2018-01-30 & 53.16 & Li \\
20947 & 2018-02-01 & 44.48 & Li \\
\hline
\end{tabular}\label{table:ChandraData}
\end{center}
\end{table}

The radio data used in this paper are the \emph{JVLA} C-array C-band observations taken from the CHANG-ES program, with detailed introduction of the program and data reduction described in \citet{Irwin12a,Irwin12b,Wiegert15}. We only use the C-array C-band data in this paper because it has high enough angular resolution to separate the radio continuum emission of the extended structures from the radio bright AGN (Walterbos et al., in prep.). The typical measurement error of the radio flux density is very low ($<1\%$). Because the major uncertainty in radio flux measurement comes from the blending of the emission from the feature of interest with other emission in the vicinity, we estimate the systematic error by adjusting the size of the flux extraction region by half a beam and calculating the resultant flux densities accordingly. The relative error calculated this way is $\approx12\%$ over the entire band.

Multi-wavelength images of the nuclear region of NGC~3079 are presented in Fig.~\ref{fig:superbubbleimg}. The superbubble on the northeast (NE) side of the galactic nucleus is best outlined with the H$\alpha$ filaments (diameter $\sim15^{\prime\prime}$), as viewed by the \emph{HST} narrow-band images first published in \citet{Cecil01}. There is also a cone-like structure capped with a few H$\alpha$ patches on the opposite side, which probably form a smaller bubble (diameter $\sim11^{\prime\prime}$). The high-resolution \emph{Chandra} soft X-ray image is highly similar to the H$\alpha$ filaments even in fine structures, strongly indicating that they are produced in related processes such as the shock heating and compression of the cool gas entrained in the outflow \citep{Cecil02}. The radio emission, however, has significantly different morphologies, with weak emission inside the H$\alpha$/soft X-ray bubble, but strong emission from a cap on top of the bubble. The radio cap is also strongly polarized, indicating a high strength for the magnetic field (e.g., \citealt{Cecil01}). Most of the emission in the hard X-ray band comes from the AGN or the nuclear starburst region and some point-like sources. We do not detect any extended hard X-ray features associated with the brighter NE bubble, but there is a cone or tuning fork like structure on the southwest (SW) side roughly coinciding with the H$\alpha$/soft X-ray filaments. Adopting a local background, we obtain a 2-7~keV detection significance of this structure to be $\rm S/N\approx6.7$ ($\rm S/N\approx3.7$ in 4-7~keV band). There may also be some hard X-ray emission associated with the cap of the SW bubble which is also visible in H$\alpha$, soft X-ray, and radio. The 2-7~keV detection significance of this ``cap'' is $\rm S/N\approx1.6$, indicating a marginal detection. The cone-like structure of the hard X-ray emission on the SW side indicates that it is unlikely produced by any detected point-like sources or scattered photons from the nuclear region.

For the convenience of discussions in the following sections, we also roughly estimate the magnetic field strength based on our new \emph{JVLA} C-array C-band data, while detailed analysis of the magnetic field will be presented elsewhere (Dettmar et al., in prep.). Assuming energy equipartition between the magnetic field and CR electrons (using the revised formula in \citealt{Beck05}), a negligible thermal contribution of $\sim1\%$ in C-band (estimated from H$\alpha$, e.g., \citealt{Vargas18}), a global non-thermal spectral index of $\alpha=0.9$, a pathlength of 1.5~kpc, the number density ratio of protons and electrons of $K_0=100$ (e.g., Stein et al., A\&A accepted), and subtracting the background halo emission of $\sim5\%$, we can obtain the magnetic field strength of the SW bubble to be $B=21.7\rm~\mu G$. We tested the robustness of the results by varying the contribution of the halo emission, the thermal fraction, and the spectral index. The resulting error of the magnetic field strength is typically $\sim1.5\rm~\mu G$. This error, however, does not include the uncertainty on the pathlength, which is hard to quantify. Similarly, we obtained a slightly higher magnetic field strength of $B=23.1\rm~\mu G$ for the NE side. A few caveats need to be made for the energy equipartition assumption adopted in the above estimate \citep{Beck15}: (1) Energy equipartition needs time to develop and hence is not valid on small length scales $\lesssim1\rm~kpc$, comparable to the size of the superbubbles in NGC~3079; (2) When $B$ varies significantly along the line of sight or within the telescope beam, the $B$ value derived from the synchrotron intensity is only a lower limit of the real field strength; (3) When energy loss of CR electrons is significant, the derived $B$ value is only a lower limit. All the above limitations may apply in NGC~3079, so the roughly estimated $B$ value can be largely uncertain (with a systematic error at least a few $\rm\mu G$, significantly larger than the estimated value of $\sim1.5\rm~\mu G$), but it is in general consistent with previous estimates using other methods (e.g., \citealt{Cecil01}).

\subsection{Spectral analysis}\label{subsec:SpecAnalysis}

We extract spectra of the \emph{Chandra} and \emph{JVLA} data from an elliptical region covering the hard X-ray cone on the SW side (called ``SW bubble'' hereafter) to examine its nature. For comparison, we also extract \emph{Chandra} spectra from an elliptical region with the same shape, area, and are symmetric with the SW bubble to the galactic nucleus (called ``NE bubble'' hereafter; Fig.~\ref{fig:superbubbleimg}). 

\begin{figure}
\begin{center}
\epsfig{figure=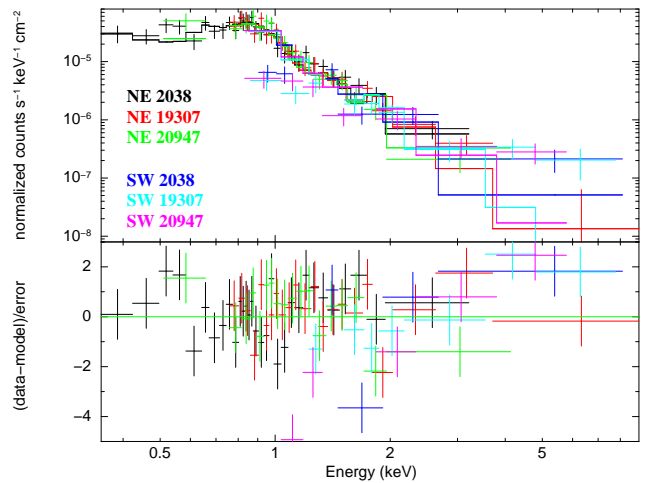,width=0.5\textwidth,angle=0, clip=}
\caption{\emph{Chandra} spectra of the NE bubble extracted from the white dashed ellipse to the NE of the nucleus of NGC~3079 (Fig.~\ref{fig:superbubbleimg}b). Different colors are spectra extracted from different observations, as denoted in the upper panel. The spectra are fitted with a single thermal plasma model (VAPEC) with the O and Fe abundances set free. The residuals at $\sim0.5-0.6\rm~keV$ are likely produced by the \ion{O}{7} lines. For comparison, we also plotted the spectra extracted from the SW bubble (blue, cyan, and magenta) which has the same area as the NE bubble. The spectra are significantly flatter than those extracted from the NE bubble. The departure from the best-fit model of the NE bubble is large, so there are a few data points of the SW bubble out of the boundary of the lower panel.}\label{fig:NEspec}
\end{center}
\end{figure}

We first analyze the X-ray spectra of the NE bubble (Fig.~\ref{fig:NEspec}), which are brighter and more extended in every band except for hard X-ray. The most prominent features of the X-ray spectra of the NE bubble are the Fe~L-shell bump at $\sim0.8-1\rm~keV$ and the \ion{O}{7} lines at $\sim0.5-0.6\rm~keV$. There may also be a weak line feature at $\sim0.6-0.7\rm~keV$ which may come from \ion{O}{8}. All these line features indicate the presence of thermal plasma, but due to the degrading of ACIS in the latest observations (ObsID 19037, 20947), we do not obtain enough photons at $\lesssim1\rm~keV$ to study the details of these emission lines. The spectra could in general be well characterized with a single thermal plasma model (VAPEC) subjected to Galactic foreground absorption ($N_{\rm H}=7.89\times10^{19}\rm~cm^{-2}$, with a best-fit upper limit of $N_{\rm H}<2.2\times10^{20}\rm~cm^{-2}$). The lack of additional extinction means the NE side is in front of the galaxy. We obtained high oxygen (${\rm Z_O/Z_{O, \odot}}=1.4_{-0.5}^{+0.6}$) and low iron (${\rm Z_{Fe}/Z_{Fe, \odot}}=0.32\pm0.05$) abundances and a temperature of $kT=0.80\pm0.04\rm~keV$ for the plasma. The total 0.5-1.5~keV luminosity of the NE bubble region is $\sim1.6\times10^{39}\rm~ergs~s^{-1}$.

Combining all the Chandra observations summarized in Table~\ref{table:ChandraData}, the 0.8-8~keV point source detection limit is $\sim3.8\times10^{38}\rm~ergs~s^{-1}$. There could be many individually faint stellar X-ray sources whose collective contribution may not be negligible (e.g., \citealt{HodgesKluck18}). We first estimate the contributions from cataclysmic variables (CVs) and coronally active binaries (ABs) whose luminosities are scaled to the stellar mass or the near-IR luminosity of the galaxies. Using the calibrated relations from \citet{Revnivtsev08}, we obtained a 0.5-1.5~keV luminosity of the CV+AB component of $\sim7\times10^{36}\rm~ergs~s^{-1}$, more than two orders of magnitude lower than that of the thermal plasma. The collected contribution from low-luminosity low-mass X-ray binaries (LMXBs) cannot be accurately estimated, because the LMXB luminosity function at the low luminosity range (typically $\lesssim10^{36-37}\rm~ergs~s^{-1}$) in star forming late-type galaxies are poorly constrained \citep{Gilfanov04}. But a rough estimate based on a detection limit of $\sim4\times10^{38}\rm~ergs~s^{-1}$ gives a 0.5-8.0~keV luminosity of the undetected LMXBs of $\sim(1-4)\times10^{38}\rm~ergs~s^{-1}$, also much lower than the thermal component. The contributions from high-mass X-ray binaries (HMXBs) is even more difficult to quantitatively estimate. This is because HMXBs are often associated with star forming regions, so the scaling relation between the total X-ray luminosity of HMXBs and the integrated SFR of the galaxy may not apply on such small scales ($\lesssim1\rm~kpc$). All the above stellar components have a significant hard X-ray excess at $\gtrsim2\rm~keV$ and the overall X-ray spectra are harder than the thermal plasma. Since we do not see a significant hard X-ray excess in the spectra, we believe the contributions from these sources are not important.

We plot the X-ray spectra of the SW bubble also in Fig.~\ref{fig:NEspec} for comparison. It is obvious that the SW bubble has much flatter X-ray spectra without much emission at $\lesssim0.8\rm~keV$. This is partially because of the degrading of ACIS in soft X-rays, while also partially because of the high extinction on this side. The counts rate at $\gtrsim4\rm~keV$ is about one order of magnitude higher than that of the NE bubble, consistent with the highly asymmetric hard X-ray morphology as revealed in the image (Fig.~\ref{fig:superbubbleimg}b). 
As the area of the NE and SW bubbles are the same and they are symmetric to the galactic center, we expect comparable contributions from the undetected stellar sources to the SW bubble as those estimated above for the NE bubble, consistent with their near-IR luminosities (only differing by $\sim15\%$). 
Therefore, these stellar source contributions at $\gtrsim2\rm~keV$ of the SW bubble should be negligible. Furthermore, the cone-shaped hard X-ray morphology of the SW bubble also rules out the possibility of a significant stellar source contribution. The flatter X-ray spectrum of the SW bubble suggests that we will need either a thermal model with a much higher temperature, or a non-thermal model to fit it. A test fit with a single thermal plasma model results in a temperature more than three times of the NE bubble, a low extinction consistent with Galactic foreground value, and residuals at both low ($\lesssim1\rm~keV$) and high ($\gtrsim4\rm~keV$) energy. These are all difficult to explain, so we will need a non-thermal component to account for the flatter hard X-ray spectrum of the SW bubble.

\begin{figure*}
\begin{center}
\epsfig{figure=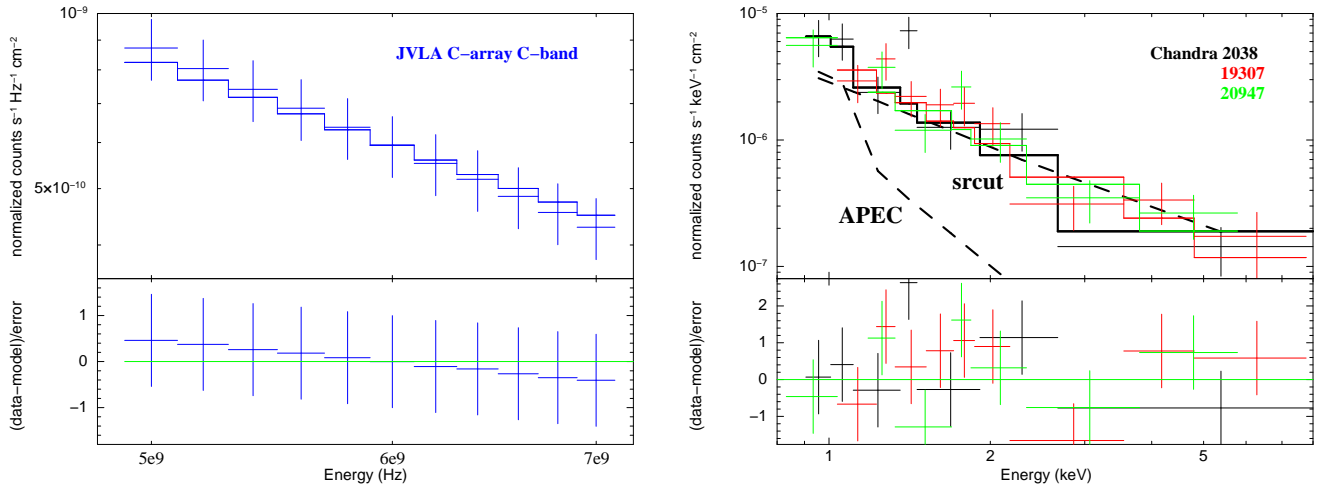,width=1.0\textwidth,angle=0, clip=}
\caption{\emph{JVLA} C-array C-band (left) and \emph{Chandra} (right) X-ray spectra of the SW bubble extracted from the white dashed ellipse to the SW of the nucleus of NGC~3079 (Fig.~\ref{fig:superbubbleimg}b). Different from X-rays, the error of the radio flux is calculated by slightly adjusting the size of the flux extraction region, as detailed in \S\ref{subsec:XrayRadioData}. The radio/X-ray spectra are jointly fitted with a ``APEC+srcut'' model with details described in the text. The two model components are plotted with dashed curves in the right panel, while for the radio spectrum, only the srcut component is important. Note that the radio and X-ray spectra are plotted with different units.}\label{fig:SWspec}
\end{center}
\end{figure*}

We jointly fit the radio/X-ray spectra of the SW bubble with an ``APEC+srcut'' model (Fig.~\ref{fig:SWspec}). The APEC component is used to account for the residual thermal contribution typically at $\sim1\rm~keV$ (very few X-ray photons are detected at $\lesssim0.8\rm~keV$), while the srcut component is used to describe the broadband synchrotron emission from radio to hard X-ray which is expected to be produced by an exponential cutoff power-law distribution of CR electrons in a homogeneous magnetic field \citep{Reynolds99}. CR electrons could also lose energy in Inverse Compton (IC) emission via the interaction with the cosmic microwave background (CMB). This IC hard X-ray emission has been detected on large scale in galaxy clusters or radio galaxies (e.g., \citealt{Isobe02,Wik09}). The relative importance of the synchrotron and IC loss depends on the energy density ratio of the magnetic field and the CMB. Typically, only when the magnetic field strength $B\lesssim3\rm~\mu G$ could the IC emission be more important (e.g., \citealt{Reynolds98}). In the nuclear region of a starburst galaxy such as NGC~3079, the strength of the magnetic field could be $B\gtrsim10\rm~\mu G$ (e.g., \citealt{Cecil01,Irwin12b}; also see \S\ref{subsec:XrayRadioData}), so the synchrotron loss should be at least one order of magnitude higher (up to $\sim50$ times higher based on the $B$ estimated in \S\ref{subsec:XrayRadioData}; the magnetic energy density $U_{\rm B}\propto B^2$). We thus believe the IC emission is less important and the broadband radio/X-ray non-thermal emission is dominated by synchrotron emission.

There may be additional intrinsic absorptions from the galaxy itself, but since we have very few photons at $\lesssim1\rm~keV$, only an upper limit of $N_{\rm H}$ can be obtained, which is: $7.5\times10^{21}\rm~cm^{-2}$. As the oxygen emission features at $<0.7\rm~keV$ are not detected, we simply assume a solar abundance of the thermal plasma and obtain a similar temperature as the NE shell of $kT=1.1\pm0.2\rm~keV$. The spectra do not show a significant cutoff. We then fix the cutoff frequency of srcut at a high value of $\nu_{\rm cutoff}=10^{25}\rm~Hz$. The exact value of $\nu_{\rm cutoff}$ will not affect our results. The best-fit spectral slope from radio to X-ray is then $\alpha_{\rm RX}=0.798_{-0.006}^{+0.018}$.

\section{Discussions and conclusions}\label{sec:discussion}

The slope of the broadband radio/X-ray spectrum ($\alpha_{\rm RX}\approx 0.8$; \S\ref{subsec:SpecAnalysis}) is slightly flatter than, or consistent within errors with, the slope determined only with the radio data in C-band ($\alpha_{\rm R}=1.11\pm0.34$; Fig.~\ref{fig:SWspec}a). Both $\alpha_{\rm RX}$ and $\alpha_{\rm R}$ may have additional systematic uncertainties which have not yet been included in the errors. The largest uncertainty may be the nature of the hard X-ray emission. Although we have discussed in \S\ref{subsec:SpecAnalysis} that it is unlikely to be a high-temperature thermal component, or faint stellar sources below the detection limit, or scattered photons from the AGN, or IC emission, most of the estimates are not quantitative enough. Furthermore, the existing X-ray spectra do not have high enough counting statistic to well decompose the non-thermal emission from other components. If the contributions from these contaminating sources are larger than expected, a steeper $\alpha_{\rm RX}$ more consistent with the measured $\alpha_{\rm R}$ may be possible.

Other contaminations may come from the radio spectra. Due to the limited angular resolution, some of the AGN emission, which is in general flatter (e.g., \citealt{Li16}), may fall in the spectral extraction region of the SW bubble. Furthermore, thermal emission in C-band may not be negligible and could contribute up to $\sim30\%$ of the flux density for some CHANG-ES galaxies \citep{Vargas18}. But both possibilities will make the real synchrotron radio spectrum even steeper, which makes the discrepancy between $\alpha_{\rm RX}$ and $\alpha_{\rm R}$ even larger. The radio spectrum may also be slightly curved, as have been revealed in many other galaxies (e.g., \citealt{Irwin12b,Wiegert15}). A positive curvature may help to explain the higher hard X-ray flux than expected from an extrapolation of the radio spectrum. A radio spectrum in a broader band is needed to better constrain its shape, especially the curvature.

Below we discuss a few problems assuming that both the radio and hard X-ray spectra are dominated by synchrotron emission. If $\alpha_{\rm RX}$ is indeed smaller than $\alpha_{\rm R}$, it means the broadband synchrotron emission flattens at higher energy. This is quite different from the commonly assumed cutoff synchrotron spectrum with the slope steepens above a characteristic energy (e.g., \citealt{Reynolds99,Li15}). Such a flattened synchrotron spectrum, however, could be produced by significant non-linear modifications of the standard diffusive shock acceleration theory (e.g., \citealt{Allen08,Caprioli12}). The flattened synchrotron spectrum and the non-linear shock acceleration scenario, however, need to be examined with better observational data.

\citet{Lacki13} examined the possible contribution of synchrotron emission to the unresolved hard X-ray emission in a few galaxies, but many uncertainties such as the stellar source contribution and magnetic field strength make the results far from conclusive. In general, because the radiative cooling timescale of 10-100~TeV CR leptons is very short, the related synchrotron hard X-ray emission is expected to be found only very close to where the CRs are accelerated. We have not only detected synchrotron hard X-ray emission on $\sim \rm kpc$ scale from the highly structured SW superbubble of NGC~3079 which is unlikely to be associated with individual star formation regions, but also have constrained the lower limit of the cutoff frequency to be $\nu_{\rm cutoff}>2\times10^{18}\rm~Hz\sim8~keV$ (without finding a significant cutoff of the broadband synchrotron spectra). This cutoff frequency is significantly higher than what we have found in smaller objects, such as SNRs (e.g., $\nu_{\rm cutoff}\sim5\times10^{16}\rm~Hz$ for the historical Galactic SNR SN1006; \citealt{Li18}). It may thus indicate collective CR accelerations on the superbubble scale and help to explain the extremely high energy CRs above the ``knee'' of the CR energy spectrum which cannot be accelerated within individual SNRs (e.g., \citealt{Kotera11}).

In the diffusive shock acceleration of CRs, the maximum energy of the CRs could be limited by a few processes, including radiative loss via synchrotron or IC emission, the finite age of the accelerator (e.g., SNR or superbubble), the maximum wavelength of MHD waves scattering the CRs to avoid the escaping of them, and the size of the accelerator compared to the gyroradius or the diffusion length of the CRs (loss-, age-, escape-, or size-limited scenarios, with the size-limited scenario often thought to be less important in SNRs; e.g., \citealt{Reynolds98,Reynolds08}). In the loss limited scenario, the maximum energy of the accelerated CRs is $E_{\rm max}\propto v_{\rm sh}B^{-1/2}$, where $v_{\rm sh}$ is the shock velocity and $B$ is the upstream magnetic field (e.g., \citealt{Reynolds08}). Compared to SNRs such as SN1006 (with an upstream $B\sim30\rm~\mu G$; e.g., \citealt{Miceli16}), we do not expect a huge difference in the $B^{-1/2}$ term for the CGM around the nuclear region of NGC~3079 ($B\sim20\rm~\mu G$, \S\ref{subsec:XrayRadioData}; \citealt{Cecil01}). On the other hand, $v_{\rm sh}$ may be a factor of $\sim10-20$ lower in the kpc-scale superbubble in NGC~3079 (typically a few hundred $\rm km~s^{-1}$ for various H$\alpha$ filaments; \citealt{Cecil01}) than those in young SNRs ($\sim2000-8000\rm~km~s^{-1}$ in SN1006, e.g., \citealt{Winkler14}). Therefore, if the radiative loss plays a dominant role in limiting the maximum energy of the accelerated CRs, we will see a much lower $\nu_{\rm cutoff}$ in the superbubble of NGC~3079 (probably in UV band), which may be consistent with the NE bubble, but definitely inconsistent with what we have detected in the SW bubble. The detection of the synchrotron hard X-ray emission in the SW bubble thus indicates that the synchrotron and IC losses must be suppressed and the maximum energy of the CRs should be limited by other mechanisms, which need to be examined with better observations and detailed modeling.

The extended diffuse hard X-ray emission in the nuclear region of NGC~3079 is highly asymmetric. The stronger hard X-ray emission in the SW bubble is very different from emissions in other bands, which are all brighter in the NE half where the extinction is smaller (Fig.~\ref{fig:superbubbleimg}). Therefore, the hard X-ray emission from the SW bubble must be intrinsic and likely coherent with the radio emission. We notice that the radio ``cap'' of the NE bubble is significantly more extended than the H$\alpha$ and soft X-ray emission, but on the SW side, they are possibly comparably vertically extended (Fig.~\ref{fig:superbubbleimg}; also see \citealt{Cecil01,Cecil02}). This may be caused by the stronger compression and amplification of the magnetic field by the apparently bigger and probably more energetic bubble on the NE side, indicating a possibly enhanced radiatively cooling of the CR electrons and thus a suppression of the hard X-ray emission. However, the estimated magnetic field strength of the NE bubble is just slightly higher than the SW side (with large uncertainty, \S\ref{subsec:XrayRadioData}) and the dependence of $E_{\rm max}$ on $B$ is weak in the loss limited scenario, so we do not expect that the synchrotron cooling should be the major reason for the asymmetry of the hard X-ray emission.

Galactic scale ($\sim\rm~kpc$ or larger) superbubbles in the nuclear region of galaxies could be produced by feedback from an AGN and/or nuclear starburst. A well known example is the ``Fermi'' bubble and its coherent multi-wavelength structures (e.g., \citealt{Finkbeiner04,Su10}), on which many models have been developed involving AGN or star formation feedback (e.g., \citealt{Guo12,Crocker15,Narayanan17}). However, because the ``Fermi'' bubble in the MW is bigger and likely older than the bubble in NGC~3079, any energetic processes happened within from a few Myrs to $\lesssim\rm Gyr$ could contribute in producing the bubble, which are difficult to track and distinguish. In contrast, assuming an expansion velocity comparable to the average velocity of the H$\alpha$ filaments (we adopt a value of $\sim300\rm~km~s^{-1}$ here; \citealt{Cecil01}), the dynamical timescale of the superbubble in NGC~3079 is only a few Myrs, comparable to the lifetime of a $25\rm~M_\odot$ star. This means the bubble must be related to the current nuclear activity. Nevertheless, it is still difficult to distinguish the AGN and starburst scenarios just based on the radio, H$\alpha$, and soft X-ray data (e.g., \citealt{Cecil02}). This is because H$\alpha$ and soft X-ray emissions are produced by shocked gas which could be produced by either AGN or starburst driven wind (e.g., \citealt{Strickland02,Guo12}), while the CR electrons responsible for the radio emission have too long cooling timescale so do not trace the in situ CR acceleration. The radio shell could be produced either by the in situ CR acceleration on the shell or by the enhanced synchrotron loss of the CRs originated from the nucleus and transported to the highly magnetized shell.

The detection of the synchrotron hard X-ray emission associated with the SW radio/H$\alpha$/soft X-ray shell is important in distinguishing the in situ acceleration and CR transport scenario. The synchrotron radiative cooling timescale $t_{\rm syn}$ in a magnetic field with strength $B$ at the peak emission frequency $\nu_{\rm m}$ can be written as (Eq. 4.54 of \citealt{You98}; also see \citealt{Lacki13} for similar discussions): $t_{\rm syn}/{\rm s}=8.7\times10^{11}(B/{\rm Gauss})^{-3/2}(\nu_{\rm m}/{\rm Hz})^{1/2}(\sin\alpha)^{-3/2}$, where $\alpha$ is the angle between the moving direction of the electron and the magnetic field. Adopting a magnetic field strength of $B\sim20\rm~\mu G$, the CR electrons responsible for the hard X-ray emission have a lifetime of only a few hundred years, far smaller than the dynamical timescale of the bubble and even significantly smaller than the diffusion timescale of the CRs from the nuclear region with any transportation mechanisms. This means that these CRs must be accelerated in situ, likely by the shock traced by the associated H$\alpha$ and soft X-ray filaments. Future deeper radio/X-ray observations, careful measurement and modeling of the magnetic field, as well as theoretical modeling of the superbubble in different scenarios, will help to better examine the nature of the hard X-ray excess in the SW bubble and to better understand the origin of galactic nuclear superbubbles.

\bigskip
\noindent\textbf{\uppercase{acknowledgements}}
\smallskip\\
\noindent The authors would like to acknowledge Rainer Beck for helpful discussions. JTL and JNB acknowledge the financial support from NASA through the grants NNX15AM93G, NAS2-97001, 80NSSC18K0536, and GO7-18073X.

\end{document}